\documentstyle[prb,aps,epsfig,floats,twocolumn]{revtex}
\begin{document}
\wideabs {
\title{
Contact Line Instability and Pattern Selection
       in Thermally Driven Liquid Films}

\author{Roman O. Grigoriev}

\address{School of Physics,
Georgia Institute of Technology, Atlanta, GA 30332-0430}

\date{\today} \maketitle

\begin{abstract} 

Liquids spreading over a solid substrate under the action of various forces are
known to exhibit a long wavelength contact line instability. We use an example
of thermally driven spreading on a horizontal surface to study how the
stability of the flow can be altered, or patterns selected, using feedback
control. We show that thermal perturbations of certain spatial structure
imposed behind the contact line and proportional to the deviation of the
contact line from its mean position can completely suppress the instability.
Due to the presence of mean flow and a spatially nonuniform nature of spreading
liquid films the dynamics of disturbances is governed by a nonnormal evolution
operator, opening up a possibility of transient amplification and nonlinear
instabilities. We show that in the case of thermal driving the nonnormality can
be significant, especially for small wavenumber disturbances, and trace the
origin of transient amplification to a close alignment of a large group of
eigenfunctions of the evolution operator. However, for values of noise likely
to occur in experiments we find that the transient amplification is not
sufficiently strong to either change the predictions of the linear stability
analysis or invalidate the proposed control approach.

\end{abstract}

\pacs{PACS numbers: 47.20.Ma, 47.54.+r, 47.62.+q} 
}
\newpage 

\section{Introduction}
\label{s_introduction}

Driven spreading of liquid films is a process which occurs in numerous
industrial coating applications, such as spin-coating of hard drives, so
understanding its dynamics and learning to control it is very important. For
instance, instabilities which arise during the spreading of the liquid on the
solid substrate can lead to nonuniform coverage, adversely affecting the
quality of produced coating. Driven spreading of thin films and patterning also
have important implications for microfluidics.

Driven spreading of liquid films under the action of gravity
\cite{jarrett,troian0}, centrifugal acceleration \cite{melo}, thermocapillary
effects \cite{cazabat}, or combination thereof \cite{troian1} has been
extensively studied in the literature. Stability analysis of such flows has
attracted the most attention and the mechanism of the linear instability is now
well understood. Considerable progress has also been reached in  feedback
control of {\em flat} liquid layers \cite{bau,kelly,grigoriev}, whose dynamics
is governed by normal differential operators. The attempts to influence the
stability of spreading films have so far been limited to  non-feedback control
achieved through either imposing an externally generated counterflow
\cite{troian2} or chemically patterning the substrate \cite{troian3,kondic}.

This study represents the first theoretical treatment of the  feedback control
problem for spreading films. The spatially nonuniform nature of spreading films
and the presence of mean flow make the control problem much more difficult
compared to the case of flat stationary films, because the dynamics of
disturbances in spreading films is governed by a nonnormal evolution operator
and thus requires a more careful analysis. For instance, liquid films flowing
down an incline have been found to develop a contact line instability when the
linear analysis predicted stable evolution \cite{bertozzi}. We derive the slip
model of thermally driven spreading and use it to show that the contact line
instability can be suppressed using adaptive thermal perturbations which
depend on the distortion of the contact line. (This type of feedback is chosen
because it is easiest to implement experimentally with sufficient spatial and
temporal resolution via optical means \cite{schatz}.) Although the results of
the following analysis should be applicable regardless of the driving force, we
concentrate our attention on the case of thermal driving.

The layout of the paper is as follows. The slip model of thermally driven
liquid films is derived in Section \ref{s_model} and its linear stability
analysis is conducted in Section \ref{s_stability}. The validity of linear
stability analysis and transient effects are considered in Section
\ref{s_nonlin}. Section \ref{s_control} presents the proposed algorithm for
feedback control of the contact line instability and Section \ref{s_discussion}
contains the conclusions.

\section{Slip model for thermal spreading}
\label{s_model}

\begin{figure}[t]
\centering
\mbox{\epsfig{figure=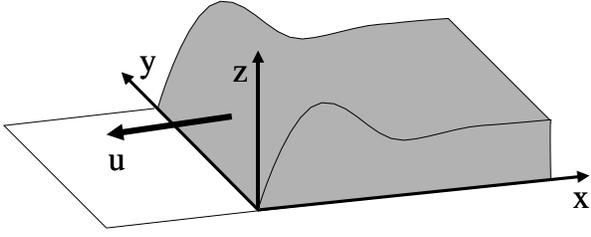,width=3.2in}}
\caption{Spreading liquid film on a solid substrate.}
\label{fig_layer}
\end{figure}

We consider the spreading of a thin layer of partially wetting liquid on a
horizontal substrate (see Fig.~\ref{fig_layer}). The spreading process is
conventionally described using the lubrication approximation \cite{davis}, with
the horizontal velocity governed by the Stokes equation
\begin{equation}
\label{eq_stokes}
\mu\partial_{zz}{\bf v}=\nabla\bar{p},
\end{equation}
where $\mu$ is the dynamic viscosity, $\bar{p}$ is the modified pressure,
$\nabla=(\partial_x,\partial_y)$ is the 2-dimensional gradient operator, and
the vertical velocity is neglected.

It is well known \cite{davis,dussan} that the standard no-slip boundary
condition at the liquid-solid interface results in a stress singularity at the
contact line. The only approach explored in the literature for the thermally
driven case was to relieve this singularity by introducing a thin precursor
film \cite{troian1}. However, as we intend to use the position of the contact
line in our control algorithm later, the precursor model becomes inconvenient.
Instead we choose to specify a microscopic contact angle and relieve the stress
singularity by employing a partial slip boundary condition
\begin{equation}
\label{bc_stress_bot}
{\bf v}=\frac{\alpha}{3h}\partial_z{\bf v}
\end{equation}
at the bottom of the liquid layer, where $h$ is the local thickness of the
film, and $\alpha$ is the phenomenological slip coefficient. The slip boundary
condition (\ref{bc_stress_bot}) was originally introduced by Greenspan
\cite{greenspan} for modeling the unforced spreading of liquid drops and later
used to model the forced spreading of liquid films under the action of gravity
\cite{homsy}. At the free surface the standard stress balance boundary
condition
\begin{equation}
\label{bc_stress_top}
\mu\partial_z{\bf v}=\nabla\sigma
\end{equation}
applies, where $\sigma$ is the surface tension coefficient. Solving
(\ref{eq_stokes}) subject to these boundary conditions we obtain the horizontal
velocity
\begin{equation}
\label{eq_velocity}
{\bf v}=\frac{1}{\mu}z\nabla\sigma
     -\frac{1}{\mu}\left(\frac{\alpha}{3}+hz
        -\frac{1}{2}z^2\right)\nabla\bar{p}.
\end{equation}
In order to make the phenomenological boundary condition (\ref{bc_stress_bot})
consistent with the physics of the flow, in (\ref{eq_velocity}) we have dropped
an unphysical term $\alpha \nabla \sigma /3\mu h$ which diverges for vanishing
film thickness. This divergence is not necessarily a cause for alarm as the
continuous approximation underlying the Stokes equation itself breaks down in
this limit. The shear stress also becomes poorly defined for very thin films.
Our choice of the functional form for the horizontal velocity
(\ref{eq_velocity}), therefore, amounts to picking an appropriate
phenomenological model in the region where the continuous description of the
flow becomes invalid and fluctuations become important.

For a film which is sufficiently thin the hydrostatic pressure can be ignored,
so that the modified pressure is given by the normal component of the surface
tension $\bar{p}=-\sigma\kappa = -\sigma \nabla^2h$. Substituting
(\ref{eq_velocity}) into the mass conservation condition
\begin{equation}
\label{bc_mass}
\partial_th=-\int_0^h(\nabla\cdot{\bf v})dz
\end{equation}
and integrating we obtain an evolution equation for the thickness:
\begin{eqnarray}
\label{eq_height}
\partial_th=-\nabla\cdot\left[\frac{1}{2\mu}h^2\nabla\sigma+
   \frac{1}{3\mu}
   (\alpha h+h^3)\nabla(\sigma\nabla^2h)\right].
\end{eqnarray}

Now consider the situation which arises when the substrate covered by the
liquid film is subjected to a linear temperature gradient in the $x$-direction.
Assuming that the surface tension changes linearly with temperature $\theta$,
\begin{equation}
\label{eq_surf_tens}
\sigma(x)=\sigma(\theta_0)+x \partial_x\theta \partial_\theta\sigma
         \equiv\sigma_0-\tau x,
\end{equation}
and neglecting the variation in $\sigma$ in the second term of
(\ref{eq_height}), which produces subdominant contribution (see, e.g., the
discussion in Ref. \onlinecite{troian1}), we obtain
\begin{eqnarray}
\label{eq_height1}
\partial_th=\frac{\tau}{2\mu}\partial_x h^2
   &-&\frac{\sigma_0}{3\mu}\partial_x[
   (\alpha h+h^3)(\partial_{xxx}h+\partial_{xyy}h)]\nonumber\\
   &-&\frac{\sigma_0}{3\mu}\partial_y[
   (\alpha h+h^3)(\partial_{xxy}h+\partial_{yyy}h)].
\end{eqnarray}
We can absorb most parameters into the spatial and temporal scales by
introducing the nondimensional variables $t'=t/T$, $x'=x/X$, $y'=y/X$, and
$h'=h/H$. The vertical length scale $H$ is defined by the characteristic
thickness of the film and sets both the horizontal length scale 
$X=[2\sigma_0H^2/3\tau]^{1/3}$ and the time scale $T=2\mu X/\tau H$. After
defining the dimensionless slip coefficient $\alpha'=\alpha/H^2$ and dropping
the primes we obtain
\begin{eqnarray}
\label{eq_height2}
\partial_th=\partial_x h^2
   &-&\partial_x[(\alpha h+h^3)(\partial_{xxx}h+\partial_{xyy}h)]\nonumber\\
   &-&\partial_y[(\alpha h+h^3)(\partial_{xxy}h+\partial_{yyy}h)].
\end{eqnarray}
The obtained equation has the same form as the one describing gravity driven
rather than temperature driven films (see, e.g., equation (33) in Ref.
\onlinecite{homsy}), with the exception that $h^2$ in the  first term on the
right-hand-side is replaced with $\alpha h+h^3$. This similarity of the
structure of the governing equations suggests that the feedback control problem
can be treated in the same way regardless of the nature of the driving force.

The liquid spreads in the direction opposite to the temperature gradient, so
the motion of the contact line is most conveniently described in the reference
frame moving with speed $u$ towards negative $x$. In this frame the evolution
equation possesses a transversely uniform steady state solution, which gives
the asymptotic film profile for constant flux boundary conditions. Substituting
$h(x,y,t)=h_0(x+ut)$ into (\ref{eq_height2}) and integrating once we obtain
\begin{equation}
\label{eq_height4}
uh_0-h_0^2+(\alpha h_0+h_0^3)h_0'''=d,
\end{equation}
where prime indicates the differentiation with respect to the $x$ coordinate.
The constants $u$ and $d$ can be determined from the appropriate boundary
conditions. Following Spaid and Homsy \cite{homsy} we require that the film
thickness vanish at the contact line, $h_0(0)=0$, and specify the slope
$h_0'(0)=c\equiv(X/H)\tan\phi$, where $\phi$ is the microscopic dynamic
contact angle in the dimensional variables. Furthermore, the constant flux
boundary condition at the tail of the film far away from the contact line gives
$h_0(\infty)=1$ (this choice corresponds to using the dimensional tail
thickness as the vertical lengthscale, $H$) and $h_0'''(\infty)=0$. These
boundary conditions require $d=0$ and $u=1$, and consequently
\begin{equation}
\label{eq_slip}
h_0'''=\frac{h_0-1}{h_0^2+\alpha}.
\end{equation}
The solution to this equation describes the height profile of the spreading
film once the distance from the contact line to the reservoir becomes
sufficiently large. 

At this point it is appropriate to make two comments regarding the structure of
the lubrication equations produced by the slip model. First of all, from a
mathematical perspective, introduction of partial slip at the liquid-solid
interface is equivalent to a regularization procedure for a singular
differential equation. Dropping the diverging term in (\ref{eq_velocity}) is
equivalent to keeping the regularization parameter $\alpha$ only in the terms
in (\ref{eq_height}) responsible for the singular behavior of the solution at
the contact line. We could have kept all the terms in (\ref{eq_velocity}) and
(\ref{eq_height}) just as well. Neither (\ref{eq_height2}) nor (\ref{eq_slip})
would have changed.

\begin{figure}[t]
\centering
\mbox{\epsfig{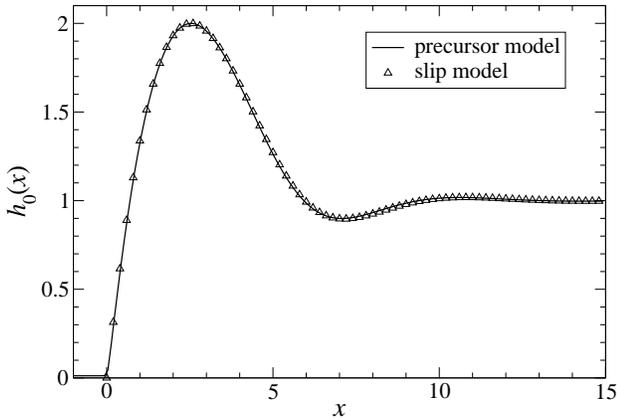}}
\vskip 3mm
\caption{Asymptotic film profiles produced by the precursor model
(\ref{eq_precursor}) with $b=0.0119$ and the slip model with $\alpha=0.1$ and
$c=1.53$. The parameters were chosen to produce a capillary ridge of height
$h_{max}=2$.}
\label{fig_pre_slip}
\end{figure}

Second, (\ref{eq_slip}) is very similar to the equation produced by the
precursor model of Kataoka and Troian \cite{troian1}
\begin{equation}
\label{eq_precursor}
h_0'''=\frac{(h_0-1)(h_0-b)}{h_0^3},
\end{equation}
where $b$ is the thickness of the precursor layer. In fact, the solutions of
the two equations are virtually indistinguishable (see Fig.~\ref{fig_pre_slip})
for the proper choice of parameters (precursor thickness, slip coefficient, and
contact angle). This is hardly a surprise, as the two equations become
identical in the limit $b\rightarrow 0$ and $\alpha \rightarrow 0$, and proves
that the final result is essentially independent of the regularization
procedure used to get rid of the stress singularity at the contact line.

\begin{figure}[t]
\centering
\mbox{\epsfig{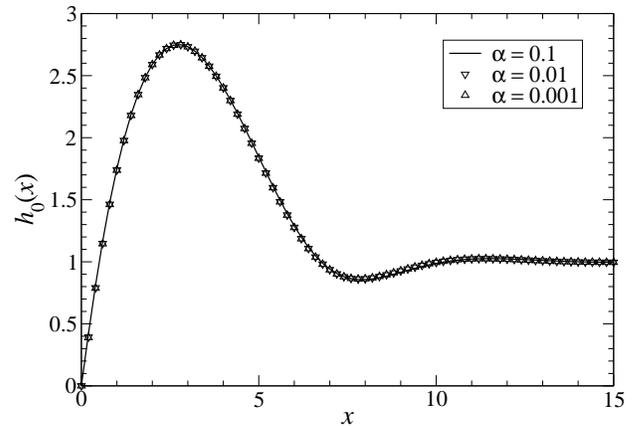}}
\vskip 3mm
\caption{Asymptotic film profiles with a fixed height of the capillary ridge
$h_{max}=2.75$ and different values of the slip coefficient.}
\label{fig_alpha}
\end{figure}

The slip model has two independent parameters ($c$ and $\alpha$), while the
precursor model has only one ($b$). A closer inspection shows that the gross
features of the slip model are, in fact, determined by a single parameter (say,
the height of the capillary ridge). As Fig.~\ref{fig_alpha} shows, one can
change the value of the slip coefficient by two orders of magnitude without
causing noticeable changes in the film profile (the contact angle will also
change in response to the changes in $\alpha$ to keep the height of the
capillary ridge constant). The second parameter fine tunes the shape of the
film in the immediate vicinity of the contact line. The slip model is,
therefore, not only more convenient for the purpose of control (it explicitly
defines the easy to measure position of the contact line), but it is to some
extent more flexible than the precursor model.

\begin{figure}[t]
\centering
\mbox{\epsfig{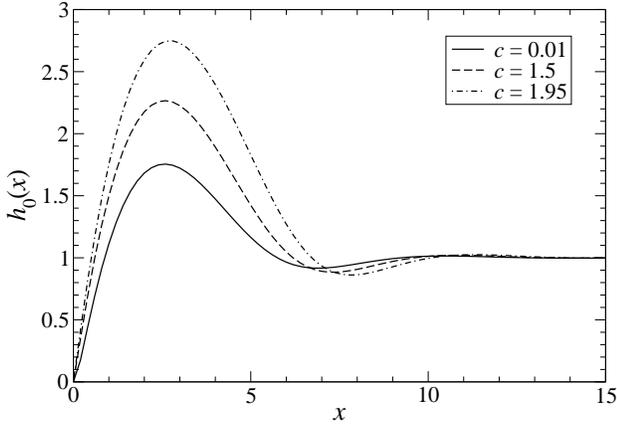}}
\vskip 3mm
\caption{Asymptotic film profiles for different values of the contact angle.
The slip coefficient is held fixed at $\alpha=0.01$.}
\label{fig_angle}
\end{figure}

Finally, let us explore the dependence of the film profile on the parameters.
As we have just seen, the profile is rather insensitive to the changes in the
slip coefficient, so let us fix it and vary the contact angle (see
Fig.~\ref{fig_angle}). The first important observation is that regardless of
the magnitude of the contact angle, the capillary ridge never disappears (this
turns out to be the case for any reasonable value of the slip coefficient,
i.e., $\alpha<1$). This is in contrast with the results obtained for gravity
driven films, where the capillary ridge completely disappears at small
inclination angles \cite{bertozzi}. Second, the height of the ridge is a
monotonically increasing function of $c$ and becomes essentially independent of
the contact angle for $c<0.3$. The dependence on the slip coefficient is more
subtle: the value of $\alpha$ controls the minimum height of the capillary
ridge which grows with decreasing slip coefficient.

\section{Contact line instability}
\label{s_stability}

Linear stability of the asymptotic solution $h_0(x+ut)$ can be determined in a
standard way\cite{troian0,troian1,bertozzi}. Since this solution is uniform in
the transverse direction, the linearized equation describing the dynamics of
small disturbances can be partially diagonalized by Fourier transforming it in
the $y$-direction. By substituting 
\begin{equation}
\label{eq_perturb1}
h(x,y,t)=h_0(x+ut)+\epsilon g(x+ut,t)e^{iqy}
\end{equation}
into (\ref{eq_height2}) and retaining terms of order $\epsilon$, we obtain
\begin{equation}
\label{eq_linear}
\partial_t g=L(q)\,g,
\end{equation}
where $L(q)= L_0+q^2L_1+q^4L_2$ is a $4^{\mathrm th}$ order differential
operator defined via
\begin{eqnarray}
\label{eq_operator012}
L_0\,g&=&-[\{1-2h_0+(\alpha+3h_0^2)h_0'''\}g+(\alpha h_0+h_0^3)g''']',
\nonumber\\
L_1\,g&=&[(\alpha+3h_0^2)h_0'g]'+2(\alpha h_0+h_0^3)g'',\nonumber\\
L_2\,g&=&-(\alpha h_0+h_0^3)g.
\end{eqnarray}
The boundary condition on $g$ at the contact line
\begin{equation}
\label{bc_head1}
h_0''(0)g(0)=h_0'(0)g'(0)
\end{equation}
can be obtained by requiring the perturbed solution to have the same contact
angle as the unperturbed solution. The other three boundary conditions can all
be imposed at the tail, say,
\begin{equation}
\label{bc_tail}
g(\infty)=g'(\infty)=g''(\infty)=0.
\end{equation}

Even though we cannot find the eigenvalues of $L(q)$ for arbitrary $q$
analytically, for long wavelength disturbances we can use perturbation theory to
get the leading order (in $q^2$) terms. This requires finding the eigenfunctions
of $L_0$ and its adjoint, $L_0^\dagger$. Taking the second derivative of 
(\ref{eq_height4}) we obtain
\begin{equation}
\label{eq_right_ef}
L_0\,h_0'=0,
\end{equation}
so that $g_0=h_0'$ is an eigenfunction of $L_0$ with eigenvalue
$\beta_0^0=0$. The adjoint operator is defined via
\begin{eqnarray}
\label{adjoint}
L_0^\dagger\,f&=&[1-2h_0+(\alpha+3h_0^2)h_0''']f'\nonumber\\
     &-&[(\alpha h_0+h_0^3)f']''',
\end{eqnarray}
so its respective eigenfunction is just a constant, say, $f_0=1$. In fact,
these are generic results with deep physical meaning. Identical relations
between the asymptotic state and the leading eigenfunctions were obtained,
e.g., for gravity driven films using the precursor model
\cite{bertozzi,kalliadasis}. The relation for $g_0$ is due to the fact that
equations for the asymptotic state are translationally invariant in the
direction of the flow (this reflects an arbitrary choice in the position of the
contact line), while the relation for $f_0$ is the consequence of the
divergence form of (\ref{bc_mass}), which reflects mass conservation.

According to the perturbation theory the leading eigenvalue has the 
following $q$-dependence:
\begin{equation}
\label{eq_eigenval1}
\beta_0(q)=\beta_0^0
+q^2\frac{\int_0^\infty f_0^*\,L_1\,g_0dx}{\int_0^\infty f_0^*g_0dx}+O(q^4).
\end{equation}
Using (\ref{eq_slip}) this can be reduced to
\begin{equation}
\label{eq_eigenval2}
\beta_0(q)=q^2\int_0^\infty h_0(h_0-1)dx+O(q^4).
\end{equation}
As this is the largest eigenvalue, its sign determines the stability of the
asymptotic state. It is easy to see that, if the asymptotic profile were
monotonic, $0<h_0<1$, the integral would be strictly negative and the system
would be linearly stable with respect to long wavelength disturbances. However,
the presence of a capillary ridge makes the integral positive, showing that the
increased mobility of the ridge provides the mechanism for the long wavelength
instability in the thermally driven case. This mechanism has been originally
conjectured by Kataoka and Troian based on the energy analysis of the precursor
model \cite{troian1}. Equation (\ref{eq_eigenval2}) gives an explicit condition
on the shape of the capillary ridge which determines the onset of instability,
and echoes a similar result obtained for the case of gravity-driven flows
\cite{troian0,bertozzi}.

Substituting $g(x,t)=h_0'(x)\exp[\beta_0(q)t]$ into (\ref{eq_perturb1}) we
notice that for small disturbances the right hand side represents the first two
terms of the Taylor expansion of a distorted asymptotic state $h_0(x+\xi+ut)$,
where
\begin{equation}
\label{eq_cont_line0}
\xi(y,t)=\epsilon e^{iqy+\beta_0(q) t}.
\end{equation}
is the deviation of the contact line from the mean. In fact, the marginal
translational mode $g_0$ is not the only eigenfunction of $L_0$. There is
an infinite discrete spectrum of eigenvalues $\beta_n$ and eigenfunctions
$g_n$. Therefore, in the presence of an arbitrary disturbance
(\ref{eq_cont_line0}) should be replaced with
\begin{equation}
\label{eq_cont_line}
\xi(y,t)=\frac{1}{c}\sum_n g_n(0)\int_{-\infty}^\infty
\epsilon_n(q)e^{iqy+\beta_n(q) t}dq.
\end{equation}

As the asymptotic state is unstable, the amplitude of the distortion will grow
exponentially in time and eventually the contact line will form equally spaced
``fingers''. In order to calculate the wavenumber of the pattern we numerically
compute the eigenfunctions and eigenvalues of the evolution operator $L(q)$.
This is most easily accomplished by discretizing the eigenvalue equation on a
spatially nonuniform mesh to properly resolve the rapid change in the solutions
near the contact line. As both the asymptotic state $h_0$ and the
eigenfunctions $g_n$ exponentially flatten for large $x$,
a truncated domain can be used, so that the boundary conditions (\ref{bc_tail})
are imposed at finite distance $l_x$ away from the contact line (we used
$l_x=80$ in most of the calculations). In order to compute the asymptotic state
we used a shooting method on a nonuniform mesh with roughly $10^4$ points.
As $h=1$ is an unstable fixed point of (\ref{eq_slip}), numerical integration
(we used the $4^{\mathrm th}$ order Runge-Kutta method) cannot be performed
beyond $l_x\approx30$ using double precision arithmetics. To determine the
solution for a longer interval we used a method \cite{tuck} in which the
numerical solution is extended using the analytical solution of (\ref{eq_slip})
for $h\approx 1$:
\begin{equation}
h(x)=1+[a\cos(\sqrt{3}\kappa x)+b\sin(\sqrt{3}\kappa x)]e^{-\kappa x}+\cdots,
\end{equation}
where $\kappa=(1+\alpha)^{-1/3}/2$ and $a$, $b$ are constants determined
by a least squares fit. The eigenvalues and eigenfunctions where then computed
using a built-in function in Matlab on a 1025-point mesh (finer resolutions did
not change the eigenvalues by more than about 5\%).

\begin{figure}[t]
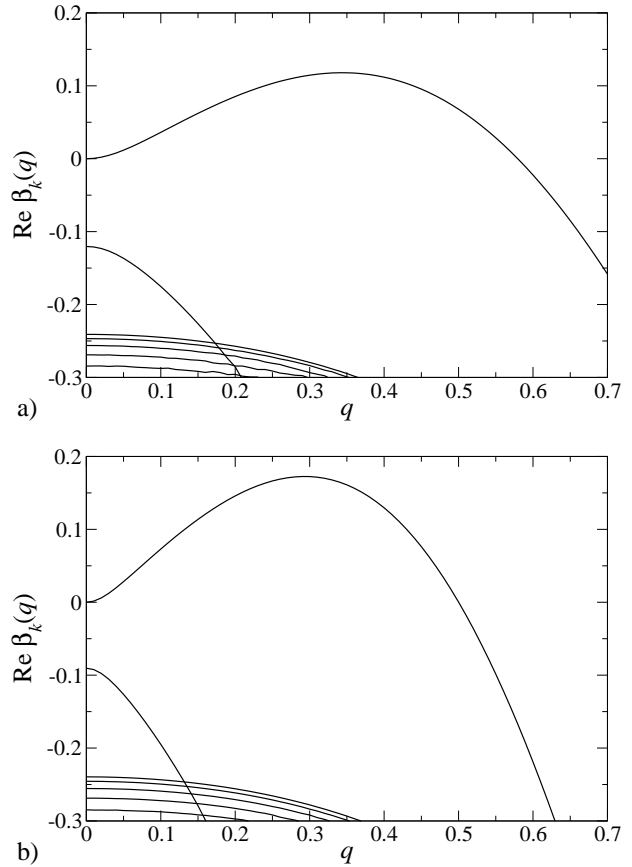

\centering
\mbox{\epsfig{figure=lambda_a.eps,width=3.2in}}
\vskip 3mm
\mbox{\epsfig{figure=lambda_b.eps,width=3.2in}}
\vskip 3mm
\caption{The twelve leading eigenvalues of $L(q)$ for $\alpha=0.01$ and (a)
$c=1$ ($h_{max}=1.93$), (b) $c=1.95$ ($h_{max}=2.75$). In both cases
$\beta_0(0)$ and $\beta_1(0)$ are real, while the other ten eigenvalues come in
complex conjugate pairs.}
\label{fig_beta}
\end{figure}

The results of our calculations for a typical choice of parameters are
presented in Fig.~\ref{fig_beta}. The fastest growing disturbance is found to
have a transverse wavenumber $q_{max}$ lying between $0.29$ and $0.34$. This
wavenumber decreases with the thickness of the capillary ridge $h_{max}$ and
gives the characteristic wavelength of the fingering pattern
$\lambda_{max}=2\pi/q_{max}$, which ranges between $18.5$ and $21.7$, in
excellent agreement with the predictions of both the precursor model
\cite{troian1} and the experiments \cite{cazabat,brzoska} which found the
dimensionless wavelength to be between $18$ and $22$. The growth rate of the
most unstable mode increases with $h_{max}$, varying between $0.12$ and $0.17$
for the range of contact angles considered here. Brzoska {\em et
al.}\cite{brzoska} have obtained an experimental value of about $0.15$, which
is also consistent with the theory. As the thickness of the ridge has not been
determined in experiments, it is impossible to make a more direct comparison,
but nevertheless these results can be used to establish the ranges of
parameters relevant for experimental conditions.

\section{Linear Stability and Transient Dynamics}
\label{s_nonlin}

We have determined earlier that the capillary ridge is present for any
reasonable choice of parameters, so the asymptotic state of a thermally driven
film is always linearly unstable and the contact line instability will
inevitably set in as the asymptotic state is approached. However, as a quick
comparison of (\ref{eq_operator012}) and (\ref{adjoint}) shows, the evolution
operator $L(q)$ is nonnormal, and we have to consider the possibility that the
transient dynamics associated with nonnormality could change the predictions of
the linear stability theory concerning the growth rates of disturbances. In
fact, transient effects could be quite significant. For instance, turbulence in
channel flows arises at values of the Reynolds number well below the critical
one predicted by the linear stability analysis \cite{trefethen}. Both in the
driven films and in channel flows nonnormality arises due to a significant mean
flow, so it is natural to expect that transient behavior could be important for
driven liquid films as well. Indeed, a disagreement between theoretical and
experimental predictions of the most unstable wavelength in gravity driven
films at low angles of inclination has been attributed to transient dynamics
\cite{bertozzi}.

Linear stability analysis presented in the previous section is based on the
assumption that the nonlinear terms are negligible at all times. If the
disturbances were small and their dynamics were governed by a normal evolution
operator, this assumption would have been well justified. For instance, when
the system is stable, the eigenvalues predict both the short term and the long
term dynamics. The situation can change dramatically when the evolution
operator becomes nonnormal, as the eigenvalues become poor predictors of the
short term dynamics. Inclusion of the nonlinear terms in (\ref{eq_linear}) has
two major consequences. First of all, nonlinear terms couple the dynamics of
modes with different transverse wavenumbers \cite{kalliadasis}. Second,
nonlinear terms produce deviations from the asymptotic state which can be
transiently amplified due to the nonnormality of the linearized evolution
operator \cite{trefethen}. A combination of these two effects can lead to a
nonlinear instability which can compete with the linear instability.

There are two scenarios which could invalidate the results of linear stability
analysis. In the simplest (purely linear) scenario \cite{bertozzi,ye}, an
initial disturbance with transverse wavenumber $q_0$ and magnitude
$\|g(x,0)\|=\xi$ could be transiently amplified by a factor $\gamma(q_0)$ to
produce a disturbance with magnitude $\gamma\xi=O(1)$. If this amplification
occurs on a time scale shorter than $1/\beta_0(q_{max})$ and
$\xi\gtrsim\gamma^{-1}$, the transient effects will dominate and a distortion
of the contact line with the wavenumber $q_0$ rather than $q_{max}$ will
result. 

\begin{figure}[t]
\centering
\mbox{\epsfig{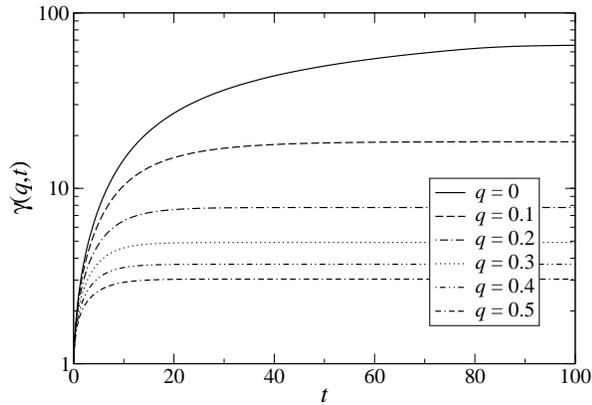}}
\vskip 3mm
\caption{Transient amplification $\gamma(q,t)$ as a function of time for
$\alpha=0.01$, $c=1$ and different values of the wavenumber.}
\label{fig_gamma_t}
\end{figure}

In a more complicated scenario \cite{trefethen,farrell}, an initial disturbance
is transiently amplified by the linear part of the evolution operator, while
the nonlinear terms produce secondary disturbances which are further
transiently amplified. This could lead to a positive feedback loop
bootstrapping a nonlinear instability, provided the secondary disturbances
contain the wavenumber of the initial disturbance and have the magnitude which
is at least as large. It is easy to check that the nonlinear evolution operator
will only contain terms quadratic, cubic, and quartic in $g(x,t)$ in addition
to the linear terms kept in (\ref{eq_linear}). Only the cubic terms will
contain the original wavenumber, so the secondary disturbances produced by the
quadratic and quartic terms will not be further transiently amplified. (Initial
disturbances with $q_0=0$ represent the only exception, but they do not lead to
distortion of the contact line.) An initial disturbance of magnitude $\xi$,
which is transiently amplified by a factor $\gamma$, will produce a secondary
disturbance of magnitude $O((\gamma\xi)^3)$ via the cubic nonlinearities. The
secondary disturbance will exceed the primary disturbance only when
$(\gamma\xi)^3\gtrsim\xi$, so that a self-sustaining nonlinear instability
becomes possible for $\xi\gtrsim\gamma^{-3/2}$.

\begin{figure}[t]
\centering
\mbox{\epsfig{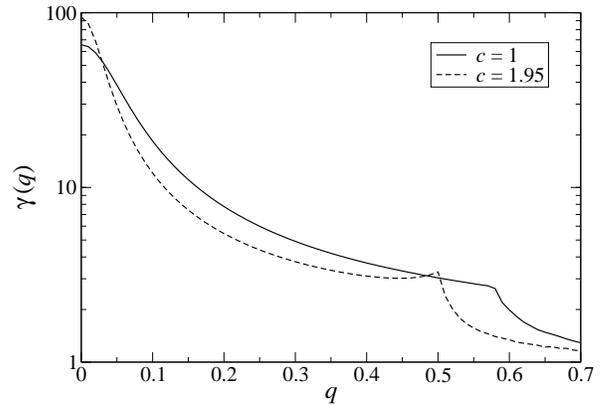}}
\vskip 3mm
\caption{Dependence of the maximal transient amplification $\gamma(q)$
on the wavenumber for $\alpha=0.01$ and different values of the contact angle.}
\label{fig_gamma_q}
\end{figure}

The maximal transient amplification in the stable band can be defined in a
conventional manner \cite{farrell}:
\begin{eqnarray}
\label{eq_gamma_s}
\gamma(q,t)=\sup_{g(x,0)}\frac{\|g(x,t)\|}{\|g(x,0)\|}
=\left\|e^{Lt}\right\|.
\end{eqnarray}
In the unstable band the fast transient amplification will be followed by the
slower exponential growth. Factoring out the exponential growth one obtains the
following upper bound:
\begin{eqnarray}
\label{eq_gamma_u}
\gamma(q,t)=\sup_{g(x,0)}\frac{\|g(x,t)\|}{\|g(x,0)\|}e^{-\beta_0 t}
=\left\|e^{(L-\beta_0)t}\right\|.
\end{eqnarray}
Numerical calculations show that in the unstable band $\gamma(q,t)$ is a
monotonically increasing function of time, so the maximum is reached for
$t\rightarrow\infty$. The time dependence  for several values of the transverse
wavenumber is shown in Fig.~\ref{fig_gamma_t}. As $L-\beta_0$ and $L$ have
the same eigenfunctions, we can easily calculate the matrix elements of the
operator $U(t)=\exp[(L-\beta_0)t]$:
\begin{equation}
U_{mn}(t)=\int_0^\infty f_m^*e^{(L-\beta_0)t}g_ndx
=e^{(\beta_n-\beta_0)t}\delta_{mn}.
\end{equation}
As only the element with $m=n=0$ survives for large times, the
maximal transient amplification is achieved for the ``optimal'' initial
disturbances equal to multiples of $f_0$. The evolution amplifies these
disturbances and transforms them into multiples of the leading eigenfunction
$g_0$. Exponential growth with rates predicted by the linear stability analysis
sets in rather quickly as the time scales of the exponential and transient
growth are of the same order of magnitude (roughly between $2$ and $9$
nondimensional units, depending on the wavenumber). 

\begin{figure}[t]
\centering
\mbox{\epsfig{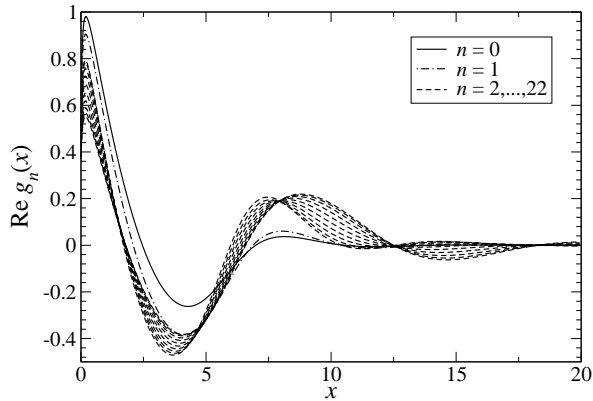}}
\vskip 3mm
\caption{The eigenfunctions $g_0(x)$ through $g_{21}(x)$ for $q=0$,
$\alpha=0.01$ and $c=1$. Only a portion of these eigenfunctions, computed on
the interval of length $l_x=80$, is shown. The eigenfunctions are normalized
such that $\int_0^\infty |g_n|^2dx=1$.}
\label{fig_right}
\end{figure}

The ultimate test of the importance of nonlinear terms and transient dynamics
is provided by a direct calculation of the transient amplification factor
\begin{equation}
\gamma(q)=\max_t\gamma(q,t).
\end{equation}
The numerical results are presented in Fig.~\ref{fig_gamma_q}. For a fixed
contact angle, $\gamma(q)$ is the largest for zero wavenumber disturbances
(which do not lead to distortion of the contact line) and quickly decreases
with increasing $q$. The maximum transient amplification increases with the
contact angle and can become quite significant for typical experimental
parameters ($\gamma(0)\approx 70$ for $c=1$, $\gamma(0)\approx 100$ for
$c=1.95$). However, the effective value for a finite system will likely be in
the range of a few tens. The strong transient amplification at $q=0$ can be
easily traced to a very close alignment of a large group of eigenfunctions
$g_2$ through $g_{21}$. As Fig.~\ref{fig_right} shows, their shapes are
extremely similar. It is, therefore, appropriate to associate the transient
behavior with a whole group of stable eigenfunctions. The size of the group
increases with $c$, increasing the degree of nonnormality. In contrast, for
channel flows apparently only a couple of near-marginal eigenfunctions become
closely aligned \cite{trefethen}.

\begin{figure}[t]
\centering
\mbox{\epsfig{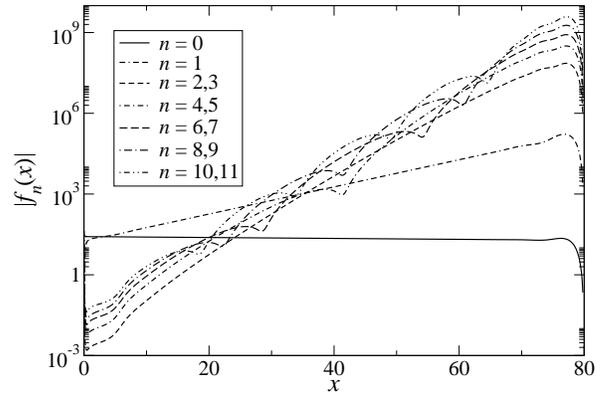}}
\vskip 3mm
\caption{The adjoint eigenfunctions $f_0(x)$ through $f_{11}(x)$ for $q=0$,
$\alpha=0.01$ and $c=1$. The leading eigenfunction $f_0(x)$ is constant, aside
from a small region near $x=l_x$, where it adjusts to the Dirichlet
boundary condition. The eigenfunctions are normalized
such that $\int_0^\infty f_n^*g_mdx=\delta_{nm}$.}
\label{fig_left}
\end{figure}

The transient amplification at small $q$ was found to depend rather sensitively
on the length of the domain used in the numerical calculations. For instance,
the values of $\gamma$ obtained for $l_x=30$ were generally about a half of
those obtained for $l_x=80$. At about $l_x=80$ the dependence levelled off, and
further increase in $l_x$ resulted in large fluctuations in the eigenvalues due
to numerical inaccuracies resulting from strong nonnormality of the matrix
produced by discretizing $L(q)$. This sensitivity can be explained by the shape
of the adjoint eigenfunctions. While the eigenfunctions $g_n(x)$ decay
exponentially for large $x$, their adjoints $f_n(x)$ {\em grow} exponentially
(see Fig.~\ref{fig_left}). As a result, small inaccuracies in the boundary
conditions at the tail of the film introduced by truncating the computational
domain can have a significant effect on the adjoint eigenfunctions, thus
affecting the transient amplification, which depends on both $g_n(x)$ and
$f_n(x)$.

The minimal noise level required to trigger the nonlinear instability according
to the second scenario is about 0.1\% of the total film thickness for
$\gamma=100$ (the first scenario requires an even higher level of noise). A
more accurate calculation of the noise threshold is likely to raise this level
much higher (to maybe a few percent) by taking into account the fact that
(\ref{eq_gamma_u}) determines the {\em maximal} amplification achieved for a
specially chosen initial condition, while the secondary disturbances will
generically be amplified less strongly. In typical experimental conditions the
noise is likely to be substantially smaller than one percent of the film
thickness. The existing experimental data \cite{cazabat,brzoska} agrees with
the predictions of the linear theory rather well, supporting the conclusion
that the transient effects are relatively weak and, therefore, the modal linear
stability analysis of the previous section accurately describes the dynamics.

\section{Feedback control of the contact line instability}
\label{s_control}

Now that we understand the limits of the linear stability analysis let us
consider the control problem. Can the contact line instability be suppressed,
or alternatively, can a pattern with a desired wavelength be imposed by
applying feedback? In principle, the answer seems to be clear, as feasibility
of feedback control of several other types of instability (buoyancy \cite{bau},
thermocapillary \cite{kelly}, evaporative \cite{grigoriev}) in liquid layers
has been demonstrated. Although one might hope that the developed control
methods could be adapted for suppressing the contact line instability, in
reality the spreading films turn out to be dramatically different.

The existing control methods have been developed for stabilizing flat films
with no mean flow, i.e., {\em steady}, {\em uniform} target states. The
evolution operators describing the dynamics of small disturbances about such
states have eigenfunctions whose horizontal dependence is given by Fourier
modes, which are normal to each other for any realistic boundary conditions (in
principle, one can {\em design} the boundary conditions such that the
eigenfunctions will not be normal even in this case \cite{handel}). Not only
does it mean that all horizontal modes become uncoupled, so any mode can be
controlled independently of the others, there is no nonnormality, so there are
no transients and the linear stability analysis is unconditionally valid.

The target state $h_0(x+ut)$ in the present problem is {\em nonuniform} in the
direction of the flow. As a result, the differential operator $L$ does not
fully diagonalize and the control problem becomes vastly more complicated. Not
only are all the modes in the system coupled, feedback applied to suppress one
mode generally affects all other modes, so an infinite-dimensional problem has
to be considered from the outset. However, even if these problems are resolved
and a feedback making the dynamics asymptotically stable is found, there is no
guarantee that the transient effects will not invalidate the whole analysis.

Let us repeat the linear stability analysis of the spreading film, but now in
the presence of feedback. First we make use of the simplification afforded by
the uniformity of the target state in the transverse direction, which allows us
to partially diagonalize the evolution operator. For the moment we restrict our
attention to monochromatic disturbances $\epsilon g(x+ut,t) \exp(iqy)$. Since
the flow is driven by the gradient in the temperature (and hence surface
tension), the stability of the flow is most easily altered by varying the
temperature field behind the contact line. Suppose we modify the temperature
profile by adding a perturbation
\begin{equation}
\label{eq_temp_contr}
\Delta\theta(x,y,t)=
-\epsilon\tau(\partial_\theta\sigma)^{-1}s(t)w(x+ut)e^{iqy},
\end{equation}
where the transverse wavelength $q$ is the same as that of the disturbance and
$s(t)$ and $w(x)$ are some functions determining the temporal and spatial
profile of the perturbation, which will be determined later. Consequently,
(\ref{eq_surf_tens}) and hence (\ref{eq_linear}) will be modified to account
for the variation in the surface tension transversely to, as well as along, the
direction of the flow. At order $\epsilon$ instead of (\ref{eq_linear}) we
obtain
\begin{eqnarray}
\label{eq_perturb}
\partial_t g&=&L_0g+s[N_1(h_0)w+N_2(h_0)w']'\nonumber\\
&+&q^2[L_1g-sN_2(h_0)w]+q^4L_2g,
\end{eqnarray}
where the influence functions
\begin{eqnarray}
\label{eq_nlc}
N_1(h_0)&=&\frac{2}{3}(h_0^2-h_0),\nonumber\\
N_2(h_0)&=&h_0^2+\frac{2}{3}(\alpha h_0+h_0^3)h_0''
\end{eqnarray}
determine the effect of the imposed thermal perturbation.

To get a sense of the dynamics of different modes in the presence of feedback,
we expand the disturbance in the basis formed by the eigenfunctions of $L_0$,
\begin{equation}
g(x,t)=\sum_m G_m(t)g_m(x),
\end{equation}
and make the strength of the applied perturbation proportional to the magnitude
of the distortion of the contact line (with a proportionality constant $k$,
called the {\em gain}, to be determined later),
\begin{equation}
s(t)=k\frac{g(0,t)}{c}=\frac{k}{c}\sum_m G_m(t)g_m(0).
\end{equation}
Multiplying (\ref{eq_perturb}) by $f_n^*$ and integrating from $0$ to
$\infty$ we obtain an infinite system of ODEs describing the dynamics of
individual modes: 
\begin{equation}
\label{eq_ode}
\dot{G}_n=\beta_n^0G_n+\sum_m (A_{nm}+q^2B_{nm}+q^4C_{nm})G_m,
\end{equation}
where (assuming that all adjoint eigenfunctions are normalized such that
$\int_0^\infty f_n^*g_mdx=\delta_{nm}$) 
\begin{eqnarray}
\label{eq_matr_elem}
A_{nm}&=&k\frac{g_m(0)}{c}\int_0^\infty
f_n^*[N_1(h_0)w+N_2(h_0)w']' dx,\nonumber\\
B_{nm}&=&\int_0^\infty f_n^*L_1g_mdx
-k\frac{g_m(0)}{c}\int_0^\infty f_n^*N_2(h_0)w\, dx,\nonumber\\
C_{nm}&=&\int_0^\infty f_n^*L_2g_m dx.
\end{eqnarray}

As Fig.~\ref{fig_beta} shows, the uncontrolled system possesses a single
unstable eigenvalue well separated from the rest of the spectrum. One can,
therefore, expect that the dynamics of small disturbances should be well
described by a single mode truncation of (\ref{eq_ode}) in the absence of
feedback. The same is not generally true when the feedback is applied. As all
modes are coupled, the feedback designed to suppress the leading mode will
always affect, and can potentially destabilize, the subleading modes. As our
numerical calculations show, such destabilization does indeed occur, unless the
spatial profile $w(x)$ of the thermal perturbation is chosen carefully to avoid
this. Had the evolution operator been {\em normal}, we could have always chosen
the feedback in such a way that different modes became uncoupled, so only the
stability of a few independent unstable modes had to be considered. The problem
of controlling the contact line instability turns out to be quite delicate in
comparison.

Assuming that $w(x)$ is chosen such that all subleading modes
remain stable, we can truncate the system (\ref{eq_ode}) by discarding all
modes except the leading one. The stability of the single-mode truncation
\begin{equation}
\label{eq_ode0}
\dot{G}_0=(A_{00}+q^2B_{00}+q^4C_{00})G_0
\end{equation}
is a necessary condition for the stability of the full system (\ref{eq_ode}).
The truncated system is stable when the following three conditions are
satisfied:
\begin{eqnarray}
\label{eq_matr_elem0}
A_{00}&=&k w'(\infty)\le 0,\nonumber\\
B_{00}&=&\int_0^\infty h_0(h_0-1)dx
-k\int_0^\infty N_2(h_0)w\, dx\le 0,\nonumber\\
C_{00}&=&-\frac{1+2\alpha}{4}\le 0.
\end{eqnarray}
The first condition is satisfied for any feedback localized near the contact
line, because in this case $A_{00}=0$, while the second condition can always be
satisfied with the proper choice of the gain $k$. The third condition is
clearly satisfied as well. Since (\ref{eq_ode}) is valid for all values of $q$
for which the governing equation (\ref{eq_height2}) is valid and the choice of
$k$ in (\ref{eq_matr_elem0}) is independent of $q$, we can immediately
generalize to non-monochromatic disturbances by integrating over all $q$, such
that the feedback will be given by
\begin{equation}
\label{eq_temp_contr1}
\Delta\theta(x,y,t)=-k\tau(\partial_\theta\sigma)^{-1}w(x+ut)\xi(y,t),
\end{equation}
where $\xi(y,t)$ is the instantaneous deviation of the contact line from its
mean position.

\begin{figure}[t]
\centering
\mbox{\epsfig{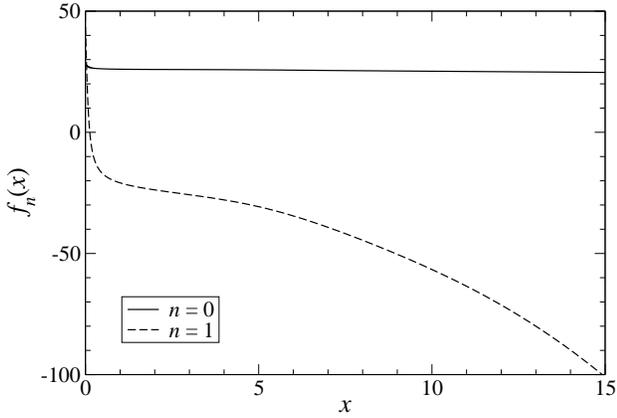}}
\vskip 3mm
\caption{The adjoint eigenfunctions $f_0(x)$ and  $f_1(x)$ for $q=0$,
$\alpha=0.01$ and $c=1$. Qualitatively similar profiles are obtained for other
values of the contact angle. A portion of the eigenvalues computed on the
interval of length $l_x=80$ is shown.}
\label{fig_adjoint}
\end{figure}

How can we choose the heating profile in the flow direction which will
stabilize the unstable mode without destabilizing any of the initially stable
modes? As Fig.~\ref{fig_right} shows, the leading mode is localized under the
capillary ridge. Therefore, to suppress the instability one has to apply a
perturbation which will be similarly localized to within the region occupied by
the ridge. The inspection of matrix elements (\ref{eq_matr_elem}) shows that
such a localized perturbation will affect all modes whose adjoint
eigenfunctions are not small in that region. According to Fig.~\ref{fig_left}
both $f_0$ and $f_1$ are of the same order of magnitude there, while the other
subleading modes are several orders of magnitude smaller. As a result, only
the stability of the leading and the first subleading mode may change in
response to feedback. The numerically computed spectra of the controlled system
support this conclusion.

The stability of the two leading modes is determined by the signs and
magnitudes of the matrix elements $A_{nm}$ and $B_{nm}$ with $n,m=0,1$. A few
general comments about these matrix elements can be made based on the structure
of the eigenfunctions. As Fig.~\ref{fig_adjoint} shows, $f_0$ and $f_1$ are
both nearly constant under the ridge and have opposite signs, while $g_0(0)$
and $g_1(0)$ have the same sign. As a result, a decrease in $B_{00}$ is
necessarily accompanied by a commensurate increase in $B_{11}$. Furthermore,
since $f_0$ is constant, we have $A_{0n}=0$ for any $n$, so at $q=0$ the
eigenvalues of the controlled system are $\beta'_0(0)=0$ and $\beta'_1(0)
\approx \beta_1(0) + A_{11}$. The matrix element $A_{11}$ is generally nonzero
and changes in response to the strength of the applied feedback. An
inappropriate choice of the profile $w(x)$ or the gain constant $k$ can make
$A_{11}$ large enough to cause destabilization at long wavelengths. However,
even when $w(x)$ is chosen such that $A_{11}$ is negative (but small),
destabilization of the $n=1$ mode at short wavelenghts can occur due to the
increase in $B_{11}$, if the feedback gain is too large.

Additional insights can be gained by considering the effect of feedback from
the physical point of view. The action of feedback in the direction transverse
to the flow is described by the influence function $N_2(h_0)$. The first and
second term of this function describe the motion of the liquid under the action
of, respectively, surface forces and pressure produced by the local gradients
in the surface tension. For a convex region of the film, where $h_0''<0$, these
two effects will compete with each other. For instance, a local maximum of
surface tension will induce the flow along the surface toward that location and
the flow in the interior of the liquid layer away from that location. The first
effect will dominate for low capillary ridges (small contact angles), the
second one for high capillary ridges (large contact angles).

\begin{figure}[t]
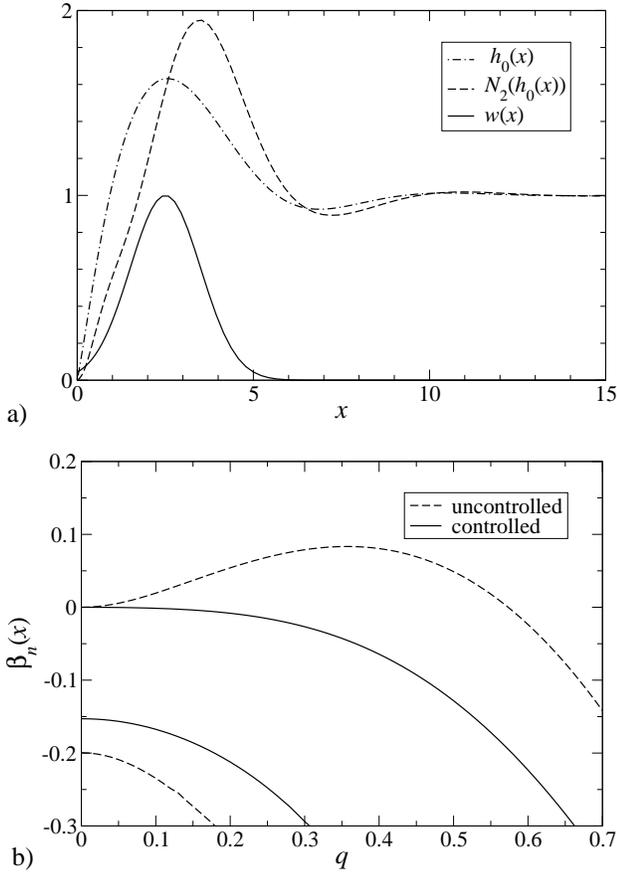

\centering
\mbox{\epsfig{figure=profile_c.eps,width=3.2in}}
\vskip 3mm
\mbox{\epsfig{figure=lam_ctl_c.eps,width=3.2in}}
\vskip 3mm
\caption{Stabilization of a thermally driven film with a low capillary ridge,
$h_{max}=1.63$ ($\alpha=0.1$ and $c=1$). (a)~The asymptotic state $h_0$,
influence function $N_2(h_0)$, and thermal perturbation profile $w$. (b)~The
two leading eigenvalues of the original and controlled system for $k=1$.}
\label{fig_low}
\end{figure}

As the instability is caused by the increased mobility of the capillary ridge, 
one might envision enforcing control by changing the thickness of the film. The
local thickness of the capillary ridge could, in principle, be modified by
locally heating or cooling it to redistribute the liquid in such a way as to
decrease the thickness, and hence the mobility, where we need to slow down the
motion of the contact line and increase the thickness and mobility, where we
need to speed it up to compensate for the deviation. This can be achieved by
choosing $w(x)$ which is localized under the ridge and does not change sign.
For instance, one could pick a Gaussian profile representing the effect of
thermal spreading in the solid substrate
\begin{equation}
\label{eq_symm}
w(x)=\exp\left[-\frac{(x-x_0)^2}{2\Delta x^2}\right],
\end{equation}
where $x_0$ and $\Delta x$ are chosen such that $w(x)$ is centered under the
capillary ridge and has a comparable width. A sample profile is shown in
Fig.~\ref{fig_low}(a) for a special choice of parameters. The numerically
computed spectrum, Fig.~\ref{fig_low}(b), demonstrates that the stabilization
can indeed be achieved by this method for driven films with a relatively low
capillary ridge.

This simple approach, however, does not always achieve the desired result. In
fact, it only succeeds when the largest growth rate $\beta_0(q_{max})$ in the
uncontrolled system is small compared to $|\beta_1(0)|$, i.e., when the
feedback required to stabilize the $n=0$ mode is too weak to destabilize the
$n=1$ mode. One can already notice the sign of approaching trouble by looking
at Fig.~\ref{fig_low}(b): the eigenvalue of the $n=1$ mode at $q=0$ starts to
creep upward due to the increase in the matrix element $A_{11}$. If this
approach is used to stabilize a flow with $\beta_0(q_{max})$ comparable to
$|\beta_1(0)|$, the feedback required to suppress the $n=0$ mode becomes strong
enough to destabilize the $n=1$ mode at low wavenumbers. The numerically
computed spectra for higher capillary ridges ($h_{max}>1.7$) show that the low
wavenumbers are destabilized before the high wavenumbers are stabilized for any
choice of the width $\Delta x$ of the thermal perturbation.  

\begin{figure}[t]
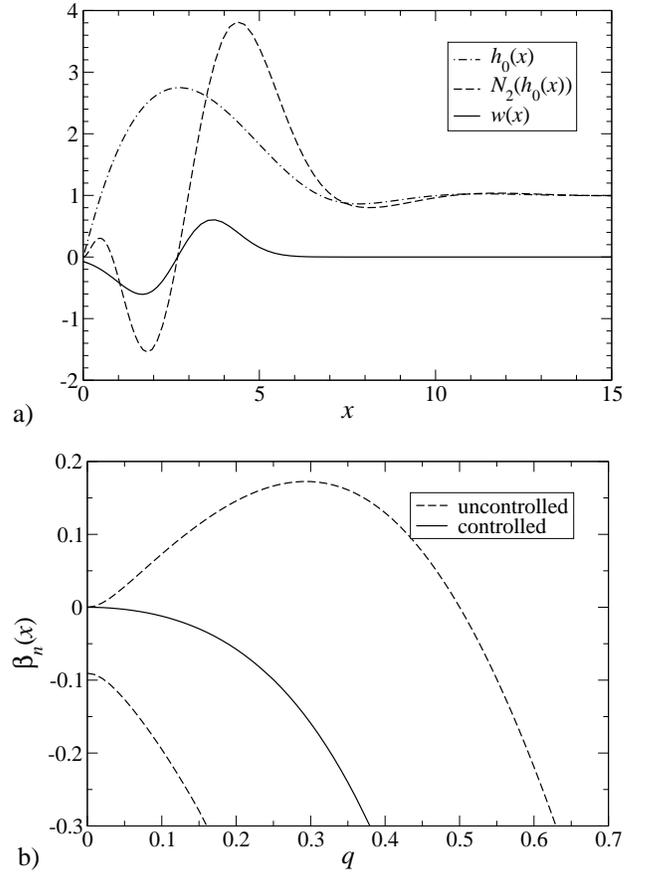

\centering
\mbox{\epsfig{figure=profile_b.eps,width=3.2in}}
\vskip 3mm
\mbox{\epsfig{figure=lam_ctl_b.eps,width=3.2in}}
\vskip 3mm
\caption{Stabilization of a thermally driven film with a high capillary ridge,
$h_{max}=2.75$ ($\alpha=0.01$ and $c=1.95$). (a)~The asymptotic state $h_0$,
influence function $N_2(h_0)$, and thermal perturbation profile $w$. (b)~The
two leading eigenvalues of the original and controlled system for $k=4$. The
$n=1$ mode is strongly suppressed by feedback, the respective eigenvalue lies
outside the graph.}
\label{fig_high}
\end{figure}

We are thus forced to look for an alternative solution. Changing the overall
thickness of the capillary ridge, even locally, is a rather ineffective
procedure, especially for small $q$, as the liquid has to be redistributed over
large distances in the transverse direction. One could instead apply a local
force to the ridge, redistributing the liquid between its front and back. For
instance, by heating the front of the ridge and cooling its back one creates
the pressure gradient enhancing the flow. Reversing the sign of the applied
perturbation impedes the flow, directly affecting the propagation velocity. The
corresponding thermal perturbation should have a profile $w(x)$ which changes
the sign near the highest point of the ridge. For instance, if one chooses the
profile to be anti-symmetric
\begin{equation}
\label{eq_asymm}
w(x)=(x-x_0)\exp\left[-\frac{(x-x_0)^2}{2\Delta x^2}\right],
\end{equation}
the matrix element $A_{11}$ can be made large and negative, so the $n=0$ mode
can be stabilized without destabilizing the $n=1$ mode. Fig.~\ref{fig_high}(a)
shows that such a thermal perturbation with the position $x_0$ and width
$\Delta x$ tuned to be roughly the same as those of the capillary ridge, should
have the largest effect on $B_{00}$. Indeed, the influence function $N_2(h_0)$,
dominated for high capillary ridges by its second term, also changes sign near
the highest point of the capillary ridge. The numerically computed spectrum,
Fig.~\ref{fig_high}(b) shows that one can again successfully suppress the
instability.

\begin{figure}[t]
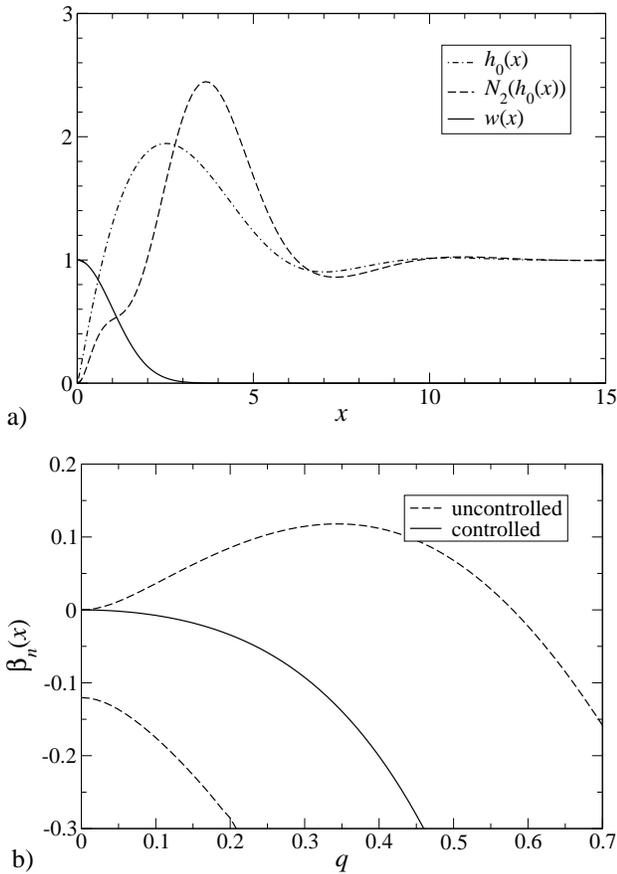

\centering
\mbox{\epsfig{figure=profile_a.eps,width=3.2in}}
\vskip 3mm
\mbox{\epsfig{figure=lam_ctl_a.eps,width=3.2in}}
\vskip 3mm
\caption{Stabilization of a thermally driven film with a medium height
capillary ridge, $h_{max}=1.93$ ($\alpha=0.01$ and $c=1$). (a)~The asymptotic
state $h_0$, influence function $N_2(h_0)$, and thermal perturbation profile
$w$. (b)~The two leading eigenvalues of the original and controlled system for
$k=4$. The $n=1$ mode is strongly suppressed by feedback, the respective
eigenvalue lies outside the graph.}
\label{fig_med}
\end{figure}

Another alternative is to exploit the narrow concave region of the film
near the contact line. One can again use the Gaussian thermal profile
(\ref{eq_symm}) centered at the contact line (see Fig.~\ref{fig_med}(a)).
Heating this region and thus lowering the surface tension one produces
gradients in both the pressure and surface tension, which induce the secondary
flow away from the contact line. Respectively, cooling this region draws the
liquid towards it, providing a direct way to locally control the propagation
speed of the film. Making the amount of heating or cooling proportional to the
displacement of the contact line again allows one to suppress the contact line
instability. The numerically computed spectrum of the system with and without
feedback is shown in Fig.~\ref{fig_med}(b).

\begin{figure}[t]
\centering
\mbox{\epsfig{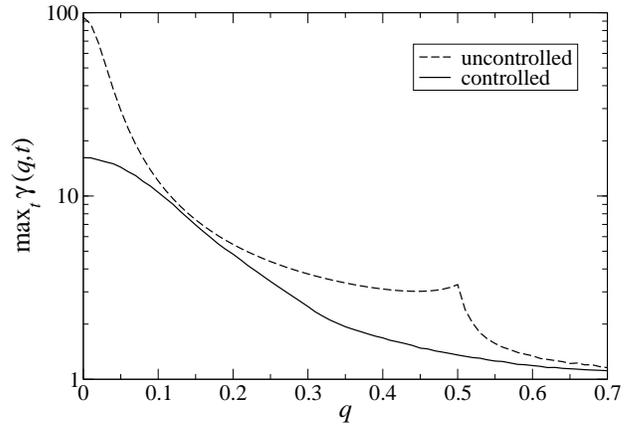}}
\vskip 3mm
\caption{The transient amplification with and without control for
$\alpha=0.01$, $c=1.95$, $k=4$, and the anti-symmetric thermal perturbation
profile (\ref{eq_asymm}).}
\label{fig_trn_ctl}
\end{figure}

Finally, let us look at the transient amplification of disturbances in the
presence of feedback. It is unclear {\em a priori} what effect the control
would have on the transient dynamics. Numerical calculations show that,
depending on the choice of the thermal profile $w(x)$, feedback can either
increase or decrease the degree of nonnormality in the system. This observation
is consistent with the theory. As the matrix elements (\ref{eq_matr_elem})
show, feedback affects different modes in a different way, and small changes in
the eigenvalues can have a large effect on the transient amplification. For
instance, thermal perturbations with an anti-symmetric profile, such as
(\ref{eq_asymm}), can decrease the transient amplification at small wavenumbers
by almost an order of magnitude (see Fig.~\ref{fig_trn_ctl}). Generally, direct
control of the propagation velocity via longitudinal surface tension gradients
leads to a decrease in the transient amplification, while indirect control via
transverse surface tension gradients affecting the mobility increases the
transient amplification for long wavelength disturbances, which is also
consistent with naive expectations.

We can thus conclude that an appropriately chosen feedback can make the
dynamics asymptotically stable without increasing the transient amplification
of disturbances, such that the contact line instability is suppressed in the
presence of noise characteristic of typical experimental conditions. Moreover,
the feedback is capable of reducing the transient amplification as well, so we
can expect that feedback control can be effective in suppressing the contact
line instability even when it is caused by the nonlinear effects. Finally, once
the instability is suppressed, selective patterning can be achieved by removing
feedback and/or introducing additional forcing at a wavenumber corresponding to
a desired pattern.

The proposed control algorithm has been verified experimentally. Although no
systematic investigation of different thermal perturbation profiles has been
attempted so far, the proof-of-principle experiments have shown that by heating
the advanced regions of the film and cooling the retarded regions one can
completely suppress the contact line instability. It has also been shown that
transversely modulated thermal perturbations applied near the contact line can
be used to achieve selective patterning, producing perfectly periodic patterns
of rivulets. The details of the experiments will be presented in a separate
publication (N. Garnier, R.~O. Grigoriev, and M.~F. Schatz, ``Optical
manipulation of microscale fluid flow'', in preparation).

\section{Discussion}
\label{s_discussion}

To summarize our results, we have determined that the evolution operator
governing the dynamics of spontaneous disturbances for thermally driven films,
both with and without feedback, is significantly nonnormal and can transiently
amplify those disturbances. The strongest transient amplification occurs for
the zero wavenumber which does not lead to contact line instability. However,
even for nonzero wavenumbers transient amplification is unlikely to produce an
instability for levels of noise characteristic of typical experimental
conditions. Therefore, linear stability analysis accurately describes both the
controlled and the uncontrolled dynamics. In contrast, for gravity driven films
at small inclination angles the transient amplification could be much stronger,
with the maximum achieved at a nonzero wavenumber \cite{bertozzi}, providing
an alternative mechanism for instability.

We have also shown that the contact line instability in thermally driven films
can be effectively suppressed by locally heating or cooling the liquid behind
the contact line. Such thermal perturbation can be easily imposed
experimentally with sufficient spatial and temporal resolution by radiatively
heating the substrate \cite{schatz}. This approach offers significant
advantages in controlling the dynamics of microflows compared to the one based
on chemical patterning of the substrate \cite{troian3,kondic}. First of all, no
preparation of the substrate is needed, while the patterns can be dynamically
reconfigured, offering potential for a significant increase in flexibility.
Second, feedback control can be used to achieve extremely small feature size,
if high intensity radiation is used to drive the flow on a thin substrate with
moderate thermal conductivity \cite{grigoriev}, opening up new prospects for
microfluidics and microfabrication applications. Finally, feedback control
provides a unique opportunity for studying the dynamics of subdominant modes
and even unstable states of the system. For instance, it can be used to
experimentally measure the growth (or decay) rates of monochromatic
disturbances with wavenumbers $q_0\ne q_{max}$ by using a wavenumber dependent
gain to suppress the disturbances which would otherwise obscure the dynamics of
the mode of interest.

We also expect that feedback control can be equally effective in suppressing
nonlinear instabilities such as those occurring in gravity driven spreading at
small inclination angles. Indeed, we have seen that the profile of the thermal
perturbations can be tuned to decrease the degree of nonnormality. Therefore,
by suppressing the transient growth feedback can also quench the bootstrapping
mechanism leading to a nonlinear instability. However, because small
disturbances at the contact line could be transiently amplified to produce
$O(1)$ changes in the thickness of the capillary ridge \cite{bertozzi}, it is
possible that the control algorithm will have to use direct measurements of the
thickness rather than the much easier to monitor position of the contact line.


\begin{references}

\bibitem{jarrett}
J. M. Jarrett and J. R. de Bruyn,``Fingering instability of a gravitationally
driven contact line,'' {\em Phys. Fluids A} {\bf 4}, 234 (1992).

\bibitem{troian0}
S. M. Troian, E. Herbolzheimer, S. A. Safran, and J. F. Joanny, ``Fingering
instabilities of driven spreading films,'' {\em Europhys. Lett.} {\bf 10}, 25
(1989).

\bibitem{melo}
F. Melo, J. F. Joanny, and S. Fauve, ``Fingering instability of spinning
drops,'' {\em Phys. Rev. Lett.} {\bf 63}, 1958 (1989).

\bibitem{cazabat}
A. M. Cazabat, F. Heslot, S. M. Troian, and P. Carles, ``Fingering instability
of thin spreading films driven by temperature gradients,'' {\em Nature} {\bf
346}, 824 (1990).

\bibitem{troian1}
D. E. Kataoka and S. M. Troian, ``A theoretical study of instabilities at the
advancing front of thermally driven coating films,'' {\em J. Colloid Interface
Sci.} {\bf 192}, 350 (1997).

\bibitem{bau}
J. Tang, H. H. Bau, ``Stabilization of the no-motion state in Rayleigh-B\'enard
convection through the use of feedback-control,'' {\it Phys. Rev. Lett.} {\bf
70}, 1795 (1993).

\bibitem{kelly}
A. C. Or, R. E. Kelly, L. Cortelezzi, and J. L. Speyer, ``Control of
long-wavelength Marangoni-B\'enard convection,'' {\em J. Fluid Mech.} {\bf
387}, 321 (1999).

\bibitem{grigoriev}
R. O. Grigoriev, ``Control of evaporatively driven instabilities of thin liquid
films,'' {\em Phys. Fluids} {\bf 14}, 1895 (2002).

\bibitem{troian2}
D. E. Kataoka and S. M. Troian, ``Stabilizing the advancing front of thermally
driven climbing films,'' {\em J. Colloid Interface Sci.} {\bf 203}, 335 (1998).

\bibitem{troian3}
D. E. Kataoka and S. M. Troian, ``Patterning liquid flow on the microscopic
scale,'' {\em Nature} {\bf 402}, 794 (1999).

\bibitem{kondic}
L. Kondic and J. Diez, ``Flow of thin films on patterned surfaces: Controlling
the instability,'' {\em Phys. Rev. E} {\bf 65}, 045301 (2002).

\bibitem{bertozzi}
A. L. Bertozzi and M. P. Brenner, ``Linear stability and transient growth in
driven contact lines,'' {\em Phys. Fluids} {\bf 9}, 530 (1997).

\bibitem{schatz}
D. Semwogerere and M. F. Schatz, ``Evolution of hexagonal patterns from
controlled initial conditions in a B\'enard-Marangoni convection experiment,''
{\em Phys. Rev. Lett.} {\bf 88}, 54501 (2002).

\bibitem{davis}
A. Oron, S. H. Davis, and S. G. Bankoff, ``Long-scale evolution of thin liquid
films,'' {\it Rev. Mod. Phys.} {\bf 69}, 931 (1997).

\bibitem{dussan}
Dussan V, S. H. Davis, ``On the motion of fluid-fluid interface along a solid
surface,'' {\em J. Fluid Mech.} {\bf 65}, 71 (1974).

\bibitem{greenspan}
H. Greenspan, ``On the motion of a small viscous droplet that wets a surface,''
{\em J. Fluid Mech.} {\bf 84}, 125 (1978).

\bibitem{homsy}
M. A. Spaid and G. M. Homsy, ``Stability of Newtonian and viscoelastic dynamic
contact lines,'' {\em Phys. Fluids} {\bf 8}, 460 (1996).

\bibitem{kalliadasis}
S. Kalliadasis, ``Nonlinear instability of a contact line driven by gravity,''
{\em J. Fluid Mech.} {\bf 413}, 355 (2000).

\bibitem{tuck}
E. O. Tuck and L. W. Schwartz, ``A numerical and asymptotic study of some
third-order ordinary differential equations relevant to draining and coating
flows,'' {\em SIAM Rev.} {\bf 32}, 453 (1990).

\bibitem{brzoska}
J. B. Brzoska, F. Brochard-Wyart, and F. Rondelez, ``Exponential-growth of
fingering instabilities of spreading films under horizontal
thermal-gradients,'' {\em Europhys. Lett.} {\bf 19}, 97 (1992).

\bibitem{trefethen} L. N. Trefethen, A. E. Trefethen, S. C. Reddy, and T. A.
Driscoll, ``Hydrodynamic stability without eigenvalues,'' {\em Science} {\bf
261}, 578 (1993).

\bibitem{ye}
Y. Ye and H.-C. Chang, ``A spectral theory for fingering on a prewetted
plane,'' {\em Phys. Fluids} {\bf 11}, 2494 (1999).

\bibitem{farrell} B. F. Farrell and P. J. Ioannou, ``Generalized stability
theory part I: autonomous operators,'' J. Atmospheric Sci. {\bf 53}, 2025
(1996).

\bibitem{handel}
R. O. Grigoriev and A. Handel, ``Nonnormality and the localized control of
extended systems,'' to be published by {\em Phys. Rev. E}.

\end{references}
\end{document}